\documentclass[12pt,english,floatfix,superscriptaddress,aps,prd,preprint,showkeys]{revtex4}
\usepackage{amsmath}
\usepackage{amssymb}
\usepackage{amsbsy}
\usepackage{amsfonts}
\usepackage{amsopn}
\usepackage{amstext}
\usepackage{graphicx}
\usepackage[english]{babel}
\usepackage{color}
\usepackage{slashed}
\usepackage{esint}
\usepackage{float}
\usepackage{units}
\usepackage{textcomp}
\usepackage{wasysym}

\usepackage{slashed}
\usepackage{hyperref}
\usepackage{graphicx}
\usepackage{amssymb}
\usepackage{amsmath}
\usepackage{color}

\begin{document}

\title{Teleparallel gravity: Effects of torsion in 6D braneworlds}

\author{A. R. P. Moreira}
\email{allan.moreira@fisica.ufc.br}
\affiliation{Universidade Federal do Cear\'a (UFC), Departamento de F\'isica,\\ Campus do Pici, Fortaleza - CE, C.P. 6030, 60455-760 - Brazil.}

\author{J. E. G. Silva}
\affiliation{Universidade Federal do Cariri(UFCA), Av. Tenente Raimundo Rocha, \\ Cidade Universit\'{a}ria, Juazeiro do Norte, Cear\'{a}, CEP 63048-080, Brasil}

\author{C.A.S. Almeida}
\email{carlos@fisica.ufc.br}
\affiliation{Universidade Federal do Cear\'a (UFC), Departamento de F\'isica,\\ Campus do Pici, Fortaleza - CE, C.P. 6030, 60455-760 - Brazil.}

\begin{abstract}
Braneworld models are interesting theoretical and phenomenological frameworks to search for new physics beyond the standard model of particles and cosmology. In this work, we discuss braneworld models whose gravitational dynamics is governed by teleparallel $f(T)$ gravities. Here, we emphasize a codimension two axisymmetric model, also known as a string-like brane. Likewise, in the 5D domain-walls models, the $f(T)$ gravitational modification leads to a phase transition on the perfect fluid source providing a brane-splitting mechanism. Furthermore, the torsion changes the gravitational perturbations. The torsion produces new potential wells inside the brane core leading to  a massless mode more localized around the ring structures. In addition, the torsion keeps a gapless non-localizable and a stable tower of massive modes in the bulk.
\end{abstract}


\maketitle

\section{Introduction}

Among the high energy theories proposed in the last years, the braneworld models gained a lot of prominence of opened new possibilities for the geometry of the multidimensional spacetime \cite{rs,rs2}, that made  possible new geometrical solutions for some of the most intriguing problems in physics, as the hierarchy problem \cite{rs2}, the origin of the dark energy \cite{Csaki1999}, dark matter \cite{darkmatter} and the origin of cosmological inflation \cite{cosmologicalconstant}. Furthermore, the warped geometry allows the bulk fields to propagate into an infinite extra dimension \cite{Csaki1,Kehagias,CASA} and provides rich internal structure for the brane \cite{Bazeia2004}. In  five dimensions, domain walls have been used to represent the brane \cite{domainwall}. In six dimensions, with axial symmetry the geometry resemblances that of the cosmic strings \cite{Gherghetta}.

The so-called modified gravity theories are models of gravity that differ from the standard Einstein General Relativity (GR) trough modifications  in curvature, namely, $f(R)$ \cite{fR01,fR1,fR2}, stress-energy tensor, $f(T)$ \cite{Yang2012} or even introducing a non-metricity feature ($f(Q)$ gravity) \cite{yo}. Among the theories of modified gravity,  $f(T)$ gravity theories have attracted much attention recently \cite{ftdarkenergy1,ftdarkenergy2,ft,Ferraroinflation, ftdegreeoffreedom, Miao,blackhole1, blackhole2, ftgw, ftpalatini, ftgw3,ftgw2, ftbinary}. Instead of metric, the dynamical variables in teleparallelism are the vielbein and,  the gravitational interaction is encoded  in the torsion \cite{Aldrovandi,Linder}.  

The $f(T)$ gravity brought interesting results  in cosmology where torsion induces changes on the stress-energy tensor \cite{ftenergyconditions}. Several works analyze the effect of the torsion in braneworlds in 5D \cite{Yang2012, Menezes,ftnoncanonicalscalar, ftborninfeld,ftmimetic,tensorperturbations}, where the $f(T)$ gravity induces changes on the stress-energy tensor. An interesting consequence is the brane splitting, since the torsion modifies the source equation of state and the dynamics of the gravitational perturbations.  

Some of the present authors, studied a $f(T)$ gravity model in a codimension-two braneworld scenario, also known as the string-like branes \cite{Moreira}. This kind of teleparallel gravity model has axially symmetric solutions of the GR  equations, which provide some interesting physical results \cite{Liu,Gherghetta, conifold,cigar,regularstring}. Inspired in the results obtained, we studied here the $f(T)$ gravity in 6D with other forms of torsion. In particular we consider an exponential dependence on the torsion as well as a hyperbolic tangent dependence one.

This work is organized as follows. In the first session we review the main definitions of the modified teleparallel theory and build the respective string-like braneworld. In this section we present the main concepts of the modified teleparallel gravity and obtain the modified gravitational equations for the braneworld scenario. Also, we find the stress energy tensor components. In the second session we derive the tensor perturbed equations and explore the gravitational Kaluza-Klein (KK) modes. Finally, additional comments are presented in the last session.

\section{Modified teleparallel braneworld}\label{sec1}

We first give a brief review of teleparallel gravity. Their spacetime metric can be constructed from the vielbein, $g_{MN}=\eta_{ab}h^a\ _M h^b\ _N$, where we assume a mostly plus metric signature $\eta_{ab}=diag(-1,1,1,1,1)$, and the capital latin index $M={0,...,D-1}$ are the bulk coordinate indexes and the latin index $a={0,...,D-1}$ is a vielbein index. In this case is used a  Weitzenb\"{o}ck connection  rather than the Levi-Civita connection that is used in general relativity. The Weitzenb\"{o}ck connection is defined as $\widetilde{\Gamma}^\rho\ _{\nu\mu}=h_a\ ^\rho\partial_\mu h^a\ _\nu$ \cite{Aldrovandi}.

The torsion tensor is constructed from the Weitzenb\"{o}ck connection, namely,  $T^{\rho}\  _{\mu\nu}= \widetilde{\Gamma}^\rho\ _{\nu\mu}-\widetilde{\Gamma}^\rho\ _{\mu\nu}$, and it is related to the Levi-Civita connection by $ K^\rho\ _{\nu\mu}=\widetilde{\Gamma}^\rho\ _{\nu\mu}-\Gamma^\rho\ _{\nu\mu}=( T_\nu\ ^\rho\ _\mu +T_\mu\ ^\rho\ _\nu - T^\rho\ _{\nu\mu})/2$. Also, it is useful to define the dual torsion tensor $S_{\rho}\ ^{\mu\nu}=( K^{\mu\nu}\ _{\rho}-\delta^\nu_\rho T^{\lambda\mu}\ _\lambda+\delta^\mu_\rho T^{\lambda\nu}\ _\lambda)/2$. Finally we have $T=T^{\rho}\  _{\mu\nu} T_{\rho}\ ^{\mu\nu}/2 +T^{\rho}\ _{\mu\nu}T^{\nu\mu}\ _{\rho}-2T^{\rho}\ _{\mu\rho}T^{\nu\mu}\ _{\nu}=T_{\rho\mu\nu}S^{\rho\mu\nu}$, which is a quadratic torsion invariant \cite{Aldrovandi}.

The $f(T)$ gravity theory is a generalization of teleparallel gravity, where the gravitational Lagrangian is a function of the torsion scalar $ T $ \cite{ft, Aldrovandi}. We assume a six dimensional bulk $f(T)$ teleparallel gravity in the form 
\begin{eqnarray}\label{55.5}
\mathcal{S}=-\frac{1}{4\kappa_g}\int h f(T)d^6x+\int h \left(\Lambda +\mathcal{L}_m\right)d^6x,
\end{eqnarray}
where $h=\sqrt{g}$, with $g$ the determinant of the metric, $\kappa_g$ is the gravitational constant and $\mathcal{L}_m$ is the matter Lagrangian. The corresponding modified gravity field equation is read as \cite{Moreira}
\begin{eqnarray}\label{3.36}
\frac{1}{h}f_T\left(\partial_Q\left(h S_N\ ^{MQ}\right)-h\widetilde{\Gamma}^R\ _{SN}S_R\ ^{MS}\right)-f_{TT}S_N\ ^{MQ}\partial_Q T& & \nonumber\\+\frac{1}{4}\delta_N^M f&=&-\kappa_g(\Lambda\delta_N^M+\mathcal{T}_N\ ^M),
\end{eqnarray}
where $f\equiv f(T)$, $f_T\equiv\partial f(T)/\partial T$ e $f_{TT}\equiv\partial^2 f(T)/\partial T^2$.

 A suitable metric ansatz for the $6D$ braneworld is given by \cite{Gherghetta,Liu}
\begin{equation}\label{45.a}
ds^2=e^{2A(r)}\eta_{\mu\nu}dx^\mu dx^\nu+dr^2+R^2_0e^{2B(r)}d\theta^2,
\end{equation}
where $0 \leq r \leq r_{max} $, $\theta \in [0; 2\pi)$, $e^{A(r)}$ and $e^{B(r)}$ are the so-called warped factors. Considering the conditions
\begin{eqnarray}
\label{regularityconditions}
e^{A(0)}=1 &,& (e^{A})'(0)=0,\nonumber\\
e^{B(0)}=0 &,& (e^{B})'(0)=1,
\end{eqnarray}
the geometry is smooth at the origin \cite{conifold,cigar}.

We then adopt \textit{sechsbeins} in the form
\begin{eqnarray}\label{0.658}
h_a\ ^M=diag(e^A, e^A, e^A, e^A, 1, R_0 e^B),
\end{eqnarray}
that give us the the torsion scalar $T=-4A'\left(3A'+2B'\right)$, where the prime $(\ '\ )$ denotes differentiation with respect to $r$. 
We assume a stress-energy tensor of form
\begin{equation}
\mathcal{T}_{N}^M=diag(t_0, t_0, t_0, t_0, t_r, t_\theta ), 
\end{equation}
due to the axisymmetric brane geometry.
Therefore, the equations of motion are given as follows
\begin{eqnarray}\label{e.1}
-\Big(6A'+2B'\Big)\Big[A''(3A'+2B')+A'(3A''+2B'')\Big]f_{TT}&&\nonumber\\ +\frac{1}{2}\Big[(4A'+B')(3A'+B')+3A''+B''\Big]f_T +\frac{1}{4}f&=&-(\Lambda+t_0),\\ 
\label{e.2}
-8A'\Big[A''(3A'+2B')+A'(3A''+2B'')\Big]f_{TT}& &\nonumber\\ + 2\Big[(4A'+B')A'+A''\Big]f_T+\frac{1}{4}f&=&-( \Lambda+t_\theta),\\ 
\label{e.3}2A'\Big(3A'+2B'\Big)f_T+\frac{1}{4}f&=&-(\Lambda+t_r),
\end{eqnarray}
where we set the gravitational constant $\kappa_g=1$ for simplicity. The equations  (\ref{e.1}), (\ref{e.2}) and (\ref{e.3}) form a quite intricate system of coupled equations, but that can be rewritten as 
\begin{eqnarray}
3A''+6A'^2+3A'B'+B'^2+B''&=&-\frac{2}{f_T}\Big(\Lambda+t_0+t_{0T} \Big),\label{0.00005}\\
2A''+5A'^2&=& -\frac{1}{f_T}\Big(\Lambda+t_\theta+t_{\theta T} \Big),\label{0.00006}\\
3A'^2+2A'B'&=& -\frac{1}{f_T}\Big(\Lambda+t_r+t_{r T} \Big),\label{0.00007}
\end{eqnarray}
where
\begin{eqnarray}
 t_{0T}&=&\frac{1}{4}f-(6A'+2B')\Big[A''(3A'+2B')+A'(3A''+2B'')\Big]f_{TT}\nonumber\\& &+A'(3A'+2B')f_T,\\
 t_{\theta T}&=&\frac{1}{4}f -8A'\Big[A''(3A'+2B')+A'(3A''+2B'')\Big]f_{TT}\nonumber\\& &+A'(3A'+2B')f_T,\\
 t_{r T}&=&\frac{1}{4}f +A'(3A'+2B')f_T.
\end{eqnarray}
Note that the left side of equations (\ref{0.00005}), (\ref{0.00006}) and (\ref{0.00007}) is equivalent to that obtained in GR. So we can states that modified gravity equations of the motion of the $f(T)$ gravities is similar to an inclusion of an additional source with $t_{0T}$, $t_{\theta T}$ and $ t_{r T}$.

The diagonal tetrad (\ref{0.658}) represent a good choice among all the possible vielbein giving metric (\ref{45.a}). In fact, the gravitational field equations do not involve any additional constraints on the function $f(T)$ or the scalar $T$. Thus, the choice in Eq.(\ref{0.658}) can be regarded as a "good vielbein" \cite{Ferraro2011,Tamanini2012}. Similarly, in a codimension one warped model the $f(T)$ gravitational dynamics preserves the form of the usual equations (two equations) \cite{Yang2012, Menezes,ftnoncanonicalscalar, ftborninfeld,ftmimetic,tensorperturbations}.

\begin{figure}
\begin{center}
\begin{tabular}{ccc}
\includegraphics[height=5cm]{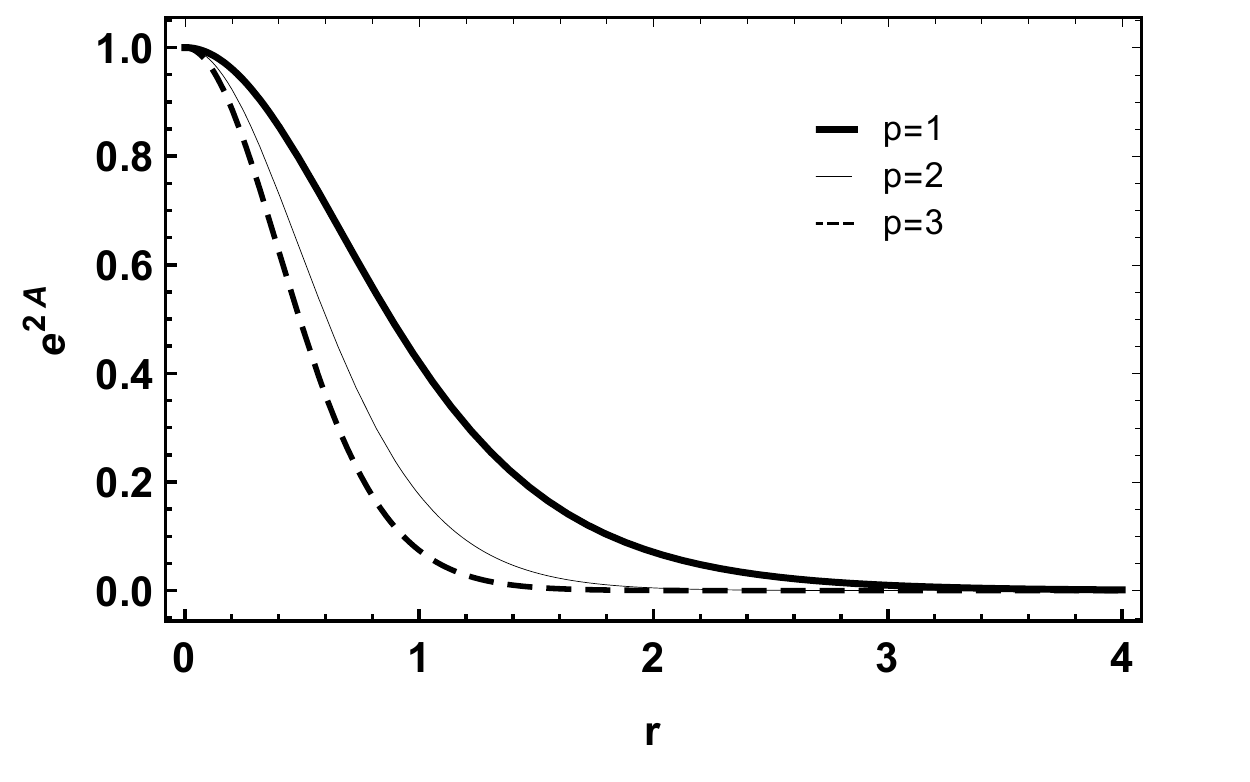}\\ 
(a)\\
\includegraphics[height=5cm]{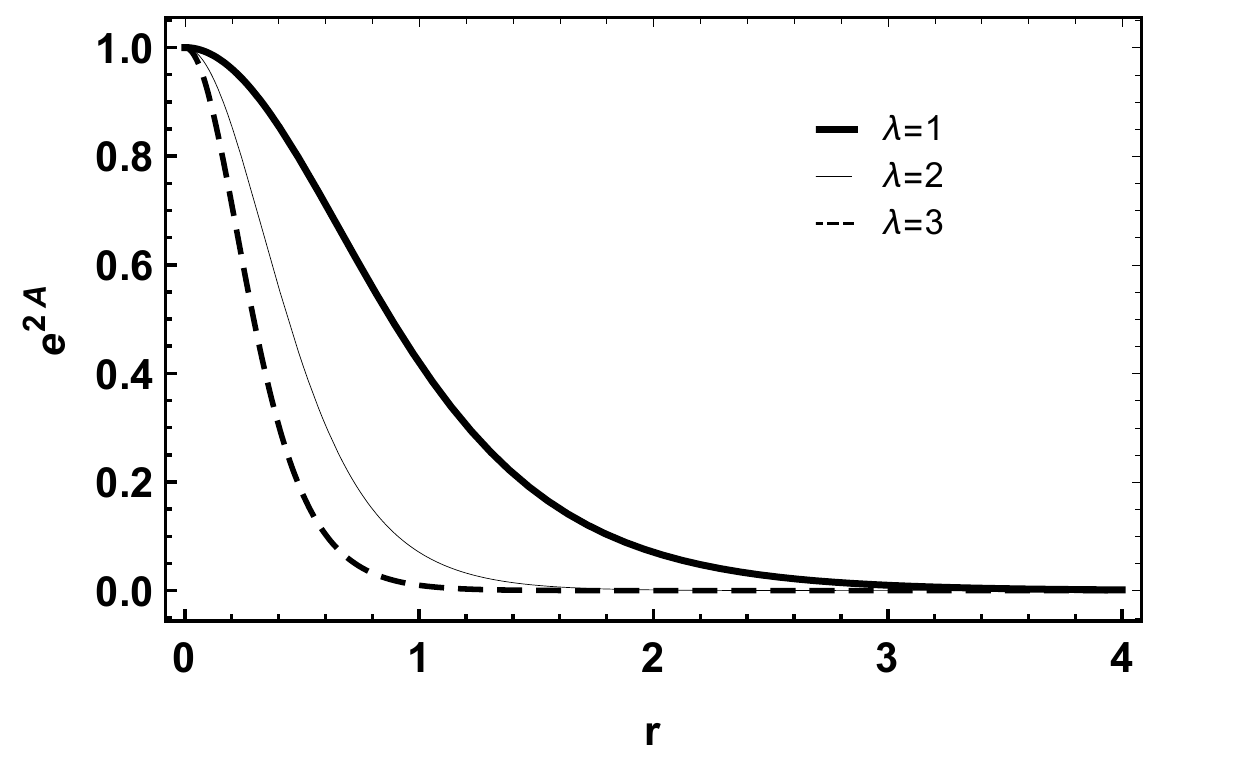}
\includegraphics[height=5cm]{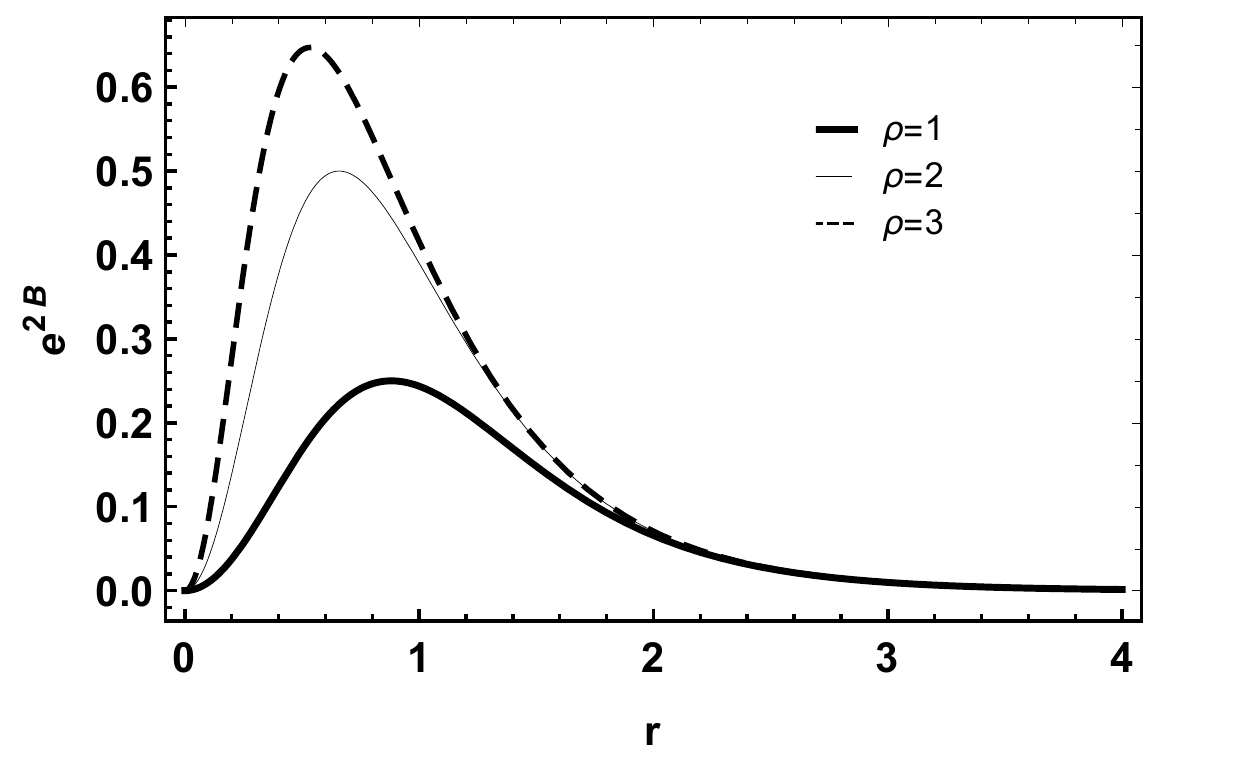}\\
(b) \hspace{8 cm}(c)
\end{tabular}
\end{center}
\caption{ Warp factor. (a) For $\lambda=1$. (b) $p=1$. (c) Angular metric component with $p=\lambda=1$.
\label{figene0}}
\end{figure}

For smooth thick string-like solutions, we propose an warp factor in the form \cite{Yang2012}
\begin{eqnarray}
\label{coreA}
e^{2A(r)}=\cosh^{-2p}(\lambda r),
\end{eqnarray}
where the parameters $p$ and $\lambda$ determine, respectively, the amplitude and the width of the source, as seen in the figure \ref{figene0} ($a$ and $b$). For the angular warp factor we assume the ansatz
\begin{eqnarray}
B(r)=\ln[\cosh^{-b}(\lambda r)]+\ln[\tanh(\rho r)],
\end{eqnarray}
where the second term guarantees the regularity condition at the origin \cite{conifold,cigar,regularstring}, its behavior can be seen in figure \ref{figene0} (c), varying the parameter $\rho$. 

Before we introduce the $f(T)$ functions used in this work, it is interesting discuss some previous important contributions from several authors that treat a torsion with an exponential dependence. First, in a 2010 paper, Eric Linder explores a modified gravity model \cite{Linder}, where a torsion tensor has an exponential dependence, in order to explain acceleration of the Universe. Soon after, a work searching for the equation of state for dark energy, uses an exponential $f(T)$ theory \cite{ftgw}. Also, some other works use $f(T)$ models with this kind of torsion, in the context of braneworlds \cite{ftnoncanonicalscalar,mirza, ftmimetic}.

The torsion is such that 
\begin{eqnarray}
T(r)=4p\lambda[4\rho\ \mathrm{csch}(2\rho r)-5p\lambda\tanh(\lambda r)]\tanh(\lambda r), 
\end{eqnarray}
and 
\begin{eqnarray}
T_0=T(0)=8p\lambda^2.
\end{eqnarray}

Let us propose first that $f_0(T)=T$, which is the case of teleparallelism. After, we will consider $f_1(T)=T_0\ e^{T/T_0}$ and $f_2(T)=T_0 \tanh({T/T_0})$.

\begin{figure}
\begin{center}
\begin{tabular}{ccc}
\includegraphics[height=5cm]{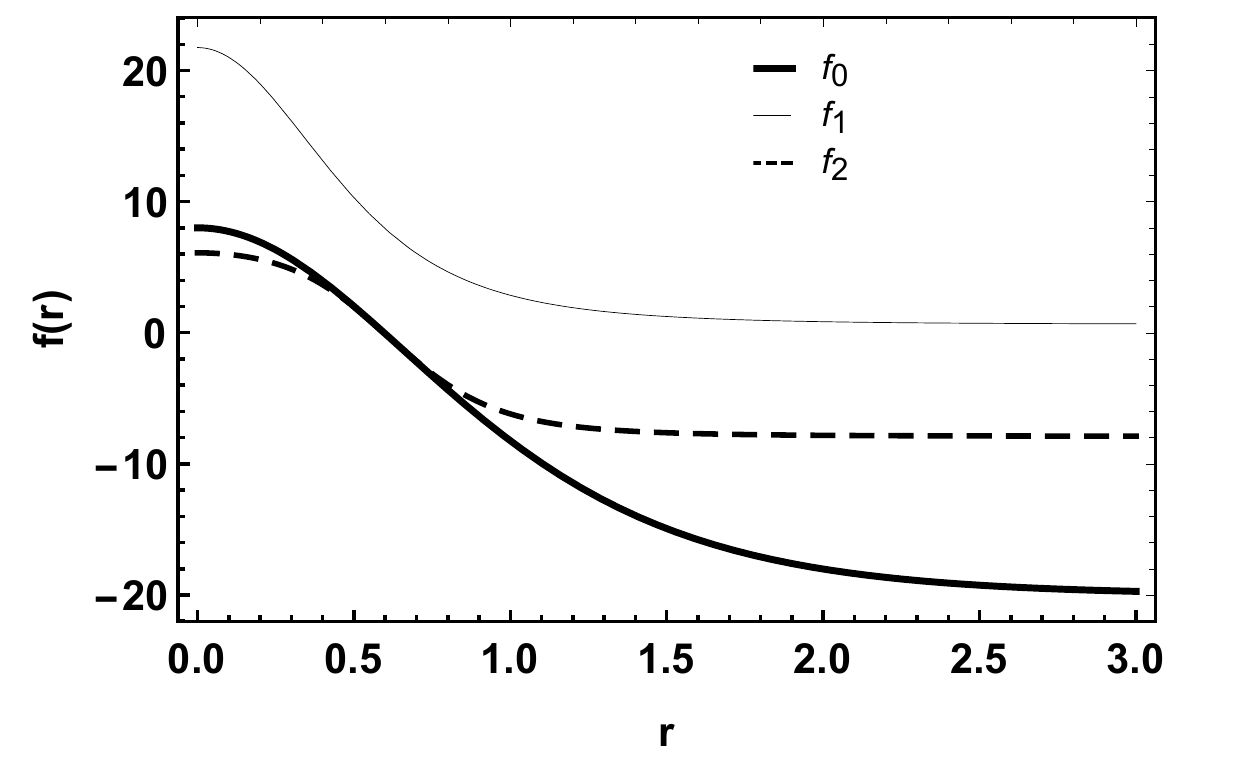}
\end{tabular}
\end{center}
\caption{$f(T)$ with $\rho=\lambda=p=1$.
\label{figf(T)1}}
\end{figure}

Now, we study the geometric features of these three cases of $f(T)$. In Fig. (\ref{figf(T)1}), we plotted the $f_0(T)$, $f_1(T)$ and $f_2(T)$ functions for this thick solution.

In order to probe further the string-like brane core, we study the distribution of the stress energy tensor components. From the modified Einstein equations (\ref{e.1}), (\ref{e.2}),\ and (\ref{e.3}), and considering the ansatz (\ref{coreA}), we get
\begin{eqnarray}
t_\theta(r)&=&-\Lambda -\frac{f}{4}+2(w+gu)f_T -\frac{16 g^2}{p\lambda}\Big\{[\mathrm{csch}(\rho r)\ \mathrm{sech}(\rho r)\rho +5]w 
+ v\Big\}f_{TT},
\end{eqnarray}
\begin{eqnarray}
t_0(r)&=&-\Lambda -\frac{f}{4}+8[2g-\rho\ \mathrm{csch}(2\rho r)]  \Big\{[5g-p\ \mathrm{csch}(\rho r)\ \mathrm{sech}(\rho r)]w +v g \Big\}f_{TT}\nonumber\\
&+&\frac{1}{2}\Big\{ [5g+9\rho\ \mathrm{sech}(\rho r)\ \mathrm{csch}(\rho r)]g+4w-u^2+v\Big\}f_T,
\end{eqnarray}
and 
\begin{eqnarray}
t_r(r)=-\Lambda -\frac{f}{4}+2g[u-2\rho\ \mathrm{csch}(2\rho r)]f_T,
\end{eqnarray}
where we define the functions $g=p \lambda\ \tanh(\lambda r)$, $w= p \lambda^2\ \mathrm{sech}^2( \lambda r)$, $v=\rho^2\mathrm{sech}^2(\rho r)+\rho^2\mathrm{csch}^2(\rho r)$, and $u=2 \rho\ \mathrm{csch}(2\rho r) -5g$.
The angular pressure vanishes asymptotically provided that for $f_0$
\begin{eqnarray}\label{e.005}
\Lambda=-5p^2\lambda^2,
\end{eqnarray}
and for $f_1$
\begin{eqnarray}\label{e.0051}
\Lambda=-2p\lambda^2 (2+10p)e^{\frac{-5p}{2}}.
\end{eqnarray}
Finally, for $f_2$
\begin{eqnarray}\label{e.005}
\Lambda=-\frac{2p\lambda^2}{(1+e^{5p})^2}\Big[e^{5p}(20p-e^{5p})+1\Big] .
\end{eqnarray}

\begin{figure}
\begin{center}
\begin{tabular}{ccc}
\includegraphics[height=5cm]{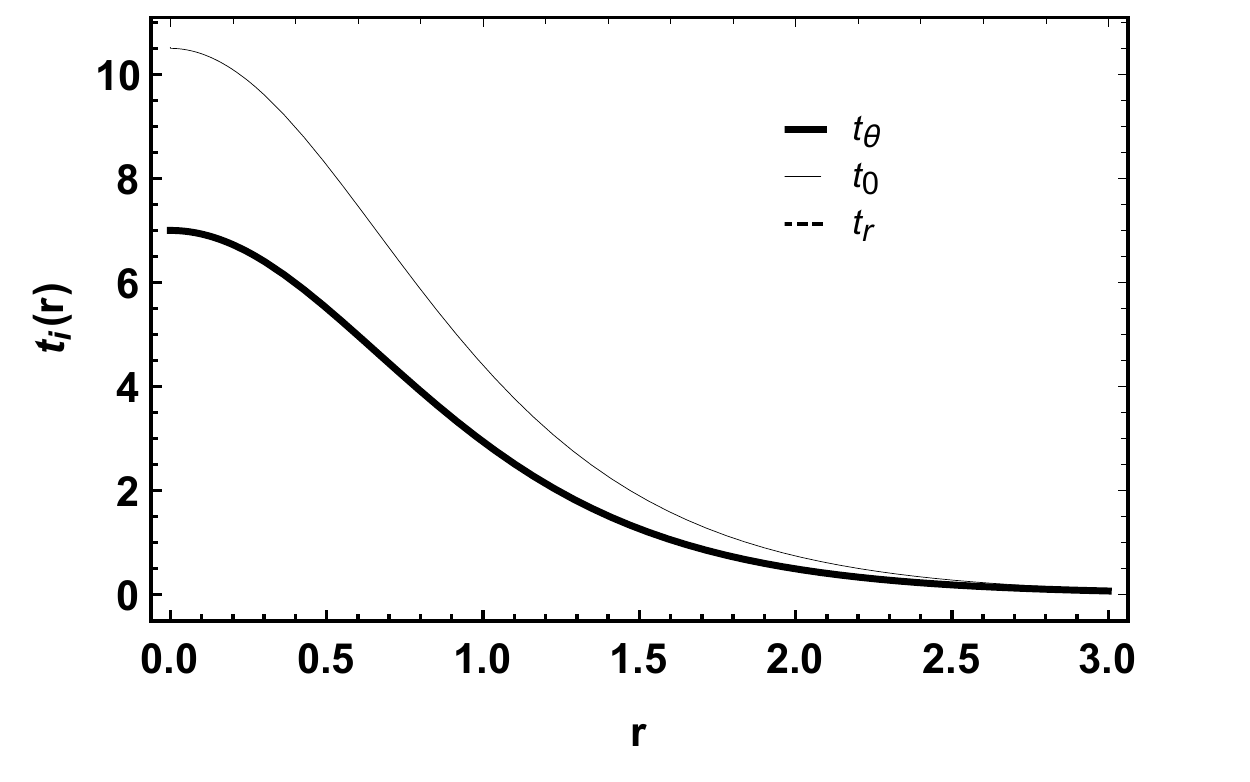}\\ 
(a)\\
\includegraphics[height=5cm]{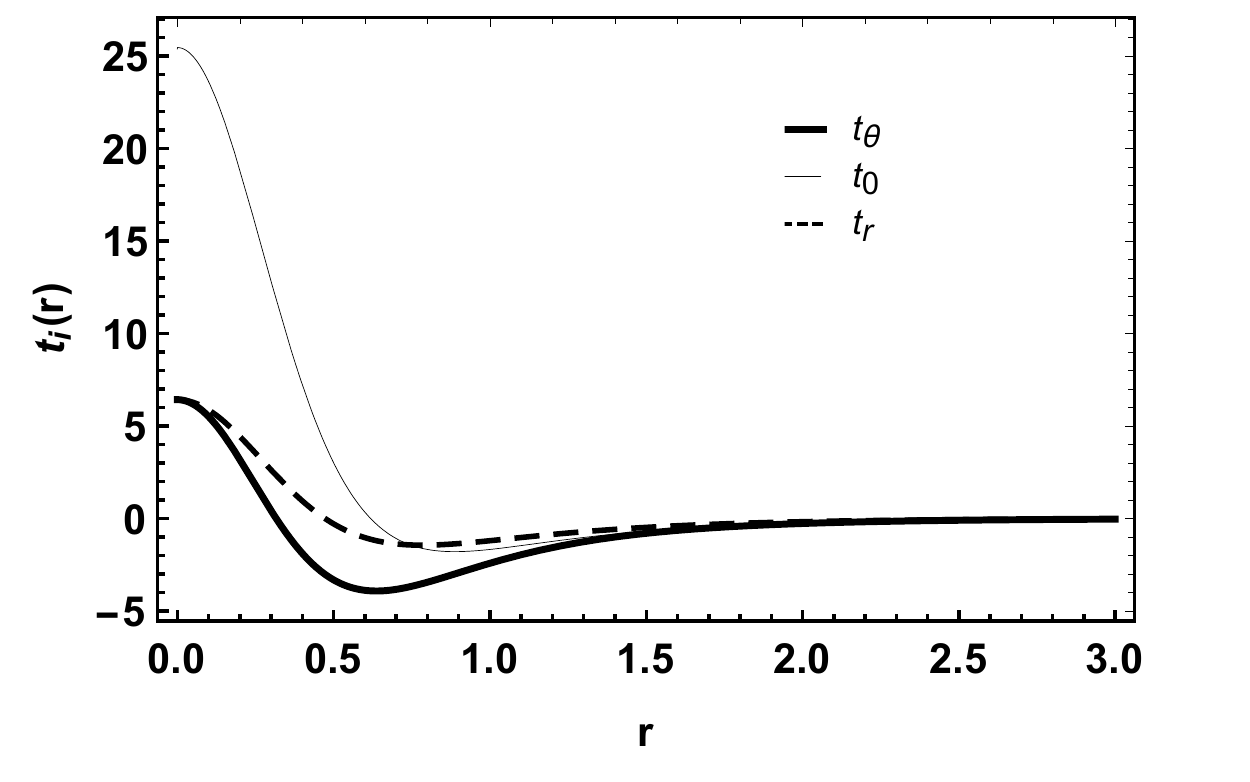}
\includegraphics[height=5cm]{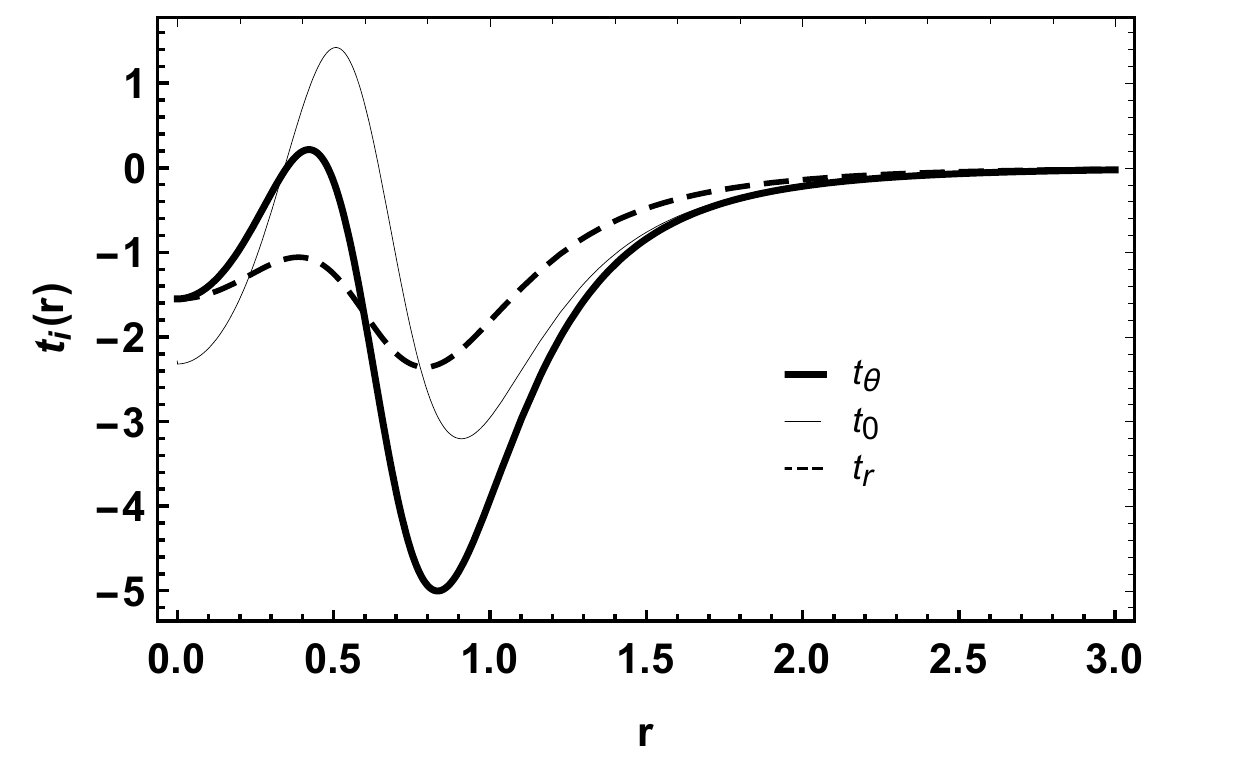}\\
(b) \hspace{8 cm}(c)
\end{tabular}
\end{center}
\caption{ Stress-energy components for  $p=\rho=\lambda=1$. (a) $f_0$. (b) $f_1$. (c) $f_2$.
\label{figene1}}
\end{figure}

In Fig.  (\ref{figene1}), we plotted the stress energy components for $f_0$, $f_1$ and $f_2$. For $f_0$ (Fig. $2a$), the source exhibits a localized profile satisfying the dominant and strong energy conditions. For $f_1$ (Fig. $2b$), the stress energy shows the appearance of a well. For $f_2$ (Fig. $2c$),  the profile includes two wells and one peak around the origin. That feature reflects the brane internal structure, which shows a splitting and forms a ring-like structure. A similar result was obtained considering a power-law modified gravity for which $f(T)=T+kT^n$, where $k$ and $n$ are two torsion parameters controlling the departure of the usual teleparallel theory \cite{Moreira}. Here, a noteworthy feature is the violation of the dominant energy condition for $f_1$ and $f_2$. Therefore, the torsion produces modifications on the source equation of state that might lead to the brane splitting. For all cases $f_0$, $f_1$ and $f_2$, increasing the value of the parameters $p$, $\lambda$ and $\rho$ increases the amplitude and width of the source.

\section{Gravitational Perturbations}
\label{sec4}

We now investigate the effects of torsion on the propagation of the gravitational perturbations.
Consider the \textit{seschsbein} perturbation \cite{Moreira}
\begin{eqnarray}\label{000588}
h^a\ _M=\left(\begin{array}{cccccc}
e^{A(r)}\left(\delta^a_\mu+w^a\ _\mu\right)&0&0\\
0&1&0\\
0&0&R_0e^{B(r)}\\
\end{array}\right),
\end{eqnarray}
where $w^a\ _\mu=w^a\ _\mu(x^\mu,r,\theta)$. Using Eq. (\ref{000588}) we can easily get the corresponding metric
\begin{eqnarray}\label{000589}
g_{MN}=\left(\begin{array}{cccccc}
e^{2A(r)}(\eta_{\mu\nu}+\gamma_{\mu\nu})&0&0\\
0&1&0\\
0&0&R^2_0e^{2B(r)}\\
\end{array}\right),
\end{eqnarray}
where the metric and the \textit{sechsbein} perturbation are related by $\gamma_{\mu\nu}=(\delta^a_\mu w^b\ _\nu+\delta^b_\nu w^a\  _\mu)\eta_{ab}$. We assume the transverse traceless metric gauge $\partial_\mu \gamma^{\mu\nu}=0=\eta^{\mu\nu}\gamma_{\mu\nu}$, which leads to the \textit{sechsbein} gauge 
$\delta_a^\mu w^a\  ^\mu=0$\cite{tensorperturbations}.

Therefore, with $\delta h=0$ and $\delta T=0$, the Eq. (\ref{3.36}) after perturbation have the form
\begin{eqnarray}\label{27.l}
f_{TT}\delta S_{MN}\ ^{Q}\partial_Q T+\frac{1}{h}f_T\Big[\delta g_{NP}\partial_Q\left(h S_M\ ^{PQ}\right)+ g_{NP}\partial_Q\left(h \delta S_M\ ^{PQ}\right)& &\nonumber\\
-h\left(\delta\widetilde{\Gamma}^Q\ _{PM}S_{QN}\ ^{P}+h\widetilde{\Gamma}^Q\ _{PM}\delta S_{QN}\ ^{P}\right)\Big] +\frac{1}{4}\delta g_{MN}f&=&\delta\mathcal{T}_{MN},
\end{eqnarray}
that lead us to
\begin{eqnarray}\label{28.l}
\delta\mathcal{T}_{\mu\nu}&=&-\frac{1}{4}\Big[e^{-2A}\Box \gamma_{\mu\nu}+(4A'+B')\gamma'_{\mu\nu}+\gamma''_{\mu\nu} +R_0^{-2}e^{-2B}\partial_\theta^2\gamma_{\mu\nu}\Big]e^{2A}f_T 
\nonumber\\ &+&\frac{1}{2}\Big[(4A'+B')(3A'+B')+3A''+B''\Big]\gamma_{\mu\nu}e^{2A}f_T+\frac{1}{4}\gamma_{\mu\nu}e^{2A}f \nonumber\\
 &-&f_{TT} [2(3A'+B')\gamma_{\mu\nu}-\gamma'_{\mu\nu}] [A''(3A'+2B')+A'(3A''+2B'')]e^{2A},
\end{eqnarray}
where $\Box=\eta^{\mu\nu}\partial_\mu \partial_\nu$ and the radial and angular perturbations vanish. The perturbation of stress energy tensor writes as $\delta\mathcal{T}_{\mu\nu}=\delta(\mathcal{T}_{\mu}\ ^\mu g_{\mu\nu})=\delta(\mathcal{T}_{\mu}\ ^\mu)\eta_{\mu\nu}e^{2A}+\mathcal{T}_{\mu}\ ^\mu \gamma_{\mu\nu}e^{2A}$. So, in our case
\begin{eqnarray}\label{30.l}
\mathcal{T}_{\mu}\  ^\mu&=&- 2(3A'+B')\left[A''(3A'+2B')+A'(3A''+2B'')\right]f_{TT}\nonumber\\
&+&\frac{1}{2}\left[(4A'+B')(3A'+B')+3A''+B''\right]f_T+\frac{1}{4}f.
\end{eqnarray}
The vanishing trace $\delta(\mathcal{T}_{\mu}\ ^\mu)$  and Eq. (\ref{30.l}) gives the perturbation equation
\begin{eqnarray}\label{32.l}
[e^{-2A}\Box \gamma_{\mu\nu}+(4A'+B')\gamma'_{\mu\nu}+\gamma''_{\mu\nu}+R_0^{-2}e^{-2B}\partial_\theta^2\gamma_{\mu\nu}]f_T& &\nonumber\\ -4 [A''(3A'+2B')+A'(3A''+2B'')]\gamma'_{\mu\nu}f_{TT}&=&0.
\end{eqnarray}
We introduce the so-called Kaluza-Klein (KK) decomposition as 
\begin{equation}
\gamma_{\mu\nu}(x^\rho,r,\theta)=\epsilon_{\mu\nu}(x^\rho)\sum_{\beta=1}^{\infty} \chi(r) e^{i\beta\theta}, 
\end{equation}
and a 4D plane-wave satisfying $\left(\Box-m^2\right)\epsilon_{\mu\nu}=0$. So, the perturbed Einstein equation (\ref{32.l}) 
leads to \cite{Moreira}
\begin{eqnarray}\label{90.l}
\Big\{4A'+B'-4 [A''(3A'+2B')+A'(3A''+2B'')]\frac{f_{TT}}{f_{T}}\Big\}\chi'& &\nonumber\\ +(e^{-2A}m^2-R_0^{-2}e^{-2B}\beta^2)\chi+\chi''&=&0.
\end{eqnarray}
In the equation above, we can see that the torsion adds a new term proportional to $f_{TT}/f_{T}$ when compared to the GR based string-like braneworld \cite{conifold,cigar,Moreira}.
\subsection{Kaluza-Klein modes}

It is interesting to consider the effects of torsion in the region exterior to the brane, which can also be interpreted as representing a thin string-like brane, where $A'=B'=-c$, and the Eq. (\ref{90.l}) takes the form \cite{Moreira}
\begin{eqnarray}\label{91.l}
\chi''-5c\chi'+e^{2cr}\left(m^2-R_0^{-2}\beta^2\right)\chi=0,
\end{eqnarray}
that is the same equation of the thin string-like brane in Gherghetta-Shaposhnikov (GS) model \cite{Gherghetta}, whose solutions can be written in terms of Bessel functions as
\begin{eqnarray}\label{92.l}
\chi(r)=e^{\frac{5}{2}c r}\left[C_1 J_{\frac{5}{2}}\left(\frac{m^2-R_0^{-2}\beta^2}{c}e^{cr}\right)+C_2 Y_{\frac{5}{2}}\left(\frac{m^2-R_0^{-2}\beta^2}{c}e^{cr}\right)\right],
\end{eqnarray}
where $C_1$ and $C_2$ are constants.

Nearby the origin, we use a Taylor serie expansion \cite{cigar,Moreira}, and the Eq. (\ref {90.l}) takes the following form for $f_0$,
\begin{eqnarray}\label{KH}
\chi''(r)+\Big[\frac{1}{r}-\frac{ r}{3}(15p\lambda^2+2\rho^2)\Big]\chi'(r)-m^2\chi(x)=0.
\end{eqnarray}
In the following we use representations of the Kummer hypergeometric confluent functions, denoted here by $_1F_1(a,b,x)$. Therefore,  the solutions of Eq. (\ref{KH}) are 
\begin{eqnarray}\label{95.1}
\chi(r)=C_0\ _1F_1 \Big(\frac{3m^2}{2(15p\lambda^2+2\rho^2)},1, \frac{1}{6}r^2(15p\lambda^2+2\rho^2)\Big).
\end{eqnarray}
Repeating the steps above, an equation for $f_1$ has as finite solutions
\begin{eqnarray}\label{95.2}
\chi(r)=C_3\ _1F_1 \Big(\frac{m^2}{4(5p\lambda^2+\rho^2)},1, r^2(5p\lambda^2+\rho^2)\Big).
\end{eqnarray}
And finally, for $f_2$ the finite solutions are 
\begin{eqnarray}\label{95.3}
\chi(r)=C_4\ _1F_1 \Bigg(\frac{3m^2(1+e^2)}{\varrho},1, \frac{r^2 \varrho}{12(1+e^2)} \Bigg),
\end{eqnarray}
where $\varrho=2[(15p\lambda^2+2\rho^2)(1+e^2)-8p\lambda^2(15p\lambda^2+4\rho^2)(1-e^2)]$, and $C_0$, $C_3$ and $C_4$ are constants.
As we can see from equations (\ref{95.1}), (\ref{95.2}) and (\ref{95.3}), the torsion modifies the KK modes.

Applying the change to a conformal coordinate $z=\int{e^{-A}}dr$ and the change on the wave function $\chi(z)=e^{-\frac{1}{2}(3A+B)+\int K(z)dz}\Psi(z)$, the KK equation can be recast in a Sch\"{o}dinger-like equation \cite{Moreira}
\begin{eqnarray}\label{36.l}
\left[-\partial_z^2+U(z)\right]\Psi(z)=m^2\Psi(z),
\end{eqnarray}
where the potential is defined by $U(z)=\dot{H}+H^2+\beta^2R_0^{-2}e^{2(A-B)}$, and the dot $(\ \dot{}\ )$  denotes differentiation with respect to $z$. Also
\begin{eqnarray}
\label{kz}
 K(z)=-4e^{-2A}\Big[3\left(\dot{A}^3-\dot{A}\ddot{A}\right)+2\dot{A}^2\dot{B}-\dot{B}\ddot{A}-\dot{A}\ddot{B}\Big] \frac{f_{TT}}{f_T},
\end{eqnarray}
and $H$ is given by 
\begin{eqnarray}\label{34.l}
H=4e^{-2A}\left[3\left(\dot{A}^3-\dot{A}\ddot{A}\right)+2\dot{A}^2\dot{B}-\dot{B}\ddot{A}-\dot{A}\ddot{B}\right] \frac{f_{TT}}{f_T}+\frac{1}{2}\left(3\dot{A}+\dot{B}\right).
\end{eqnarray}
When $\beta=0$, we have the quantum mechanic supersymmetric form of the potential $U$ that ensures the absence of tachyonic KK gravitational models \cite{CASA}. Besides the spectrum stability, the potential allows a massless KK mode of form  \cite{Moreira}
\begin{eqnarray}
\Psi_0=N_0e^{\frac{1}{2}(3A+B))-\int K(z)dz},
\end{eqnarray}
where $N_0$ is a normalization constant. 

\begin{figure}
\begin{center}
\begin{tabular}{ccccccccc}
\includegraphics[height=5cm]{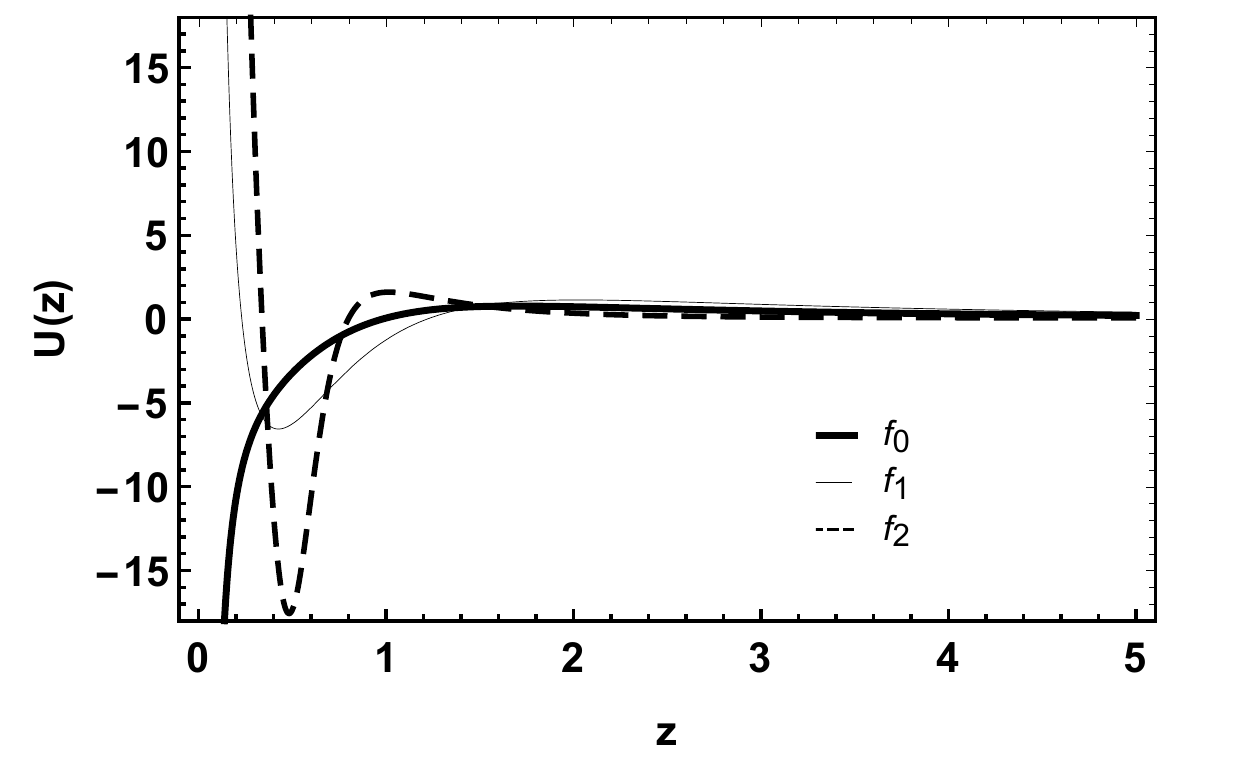}
\includegraphics[height=5cm]{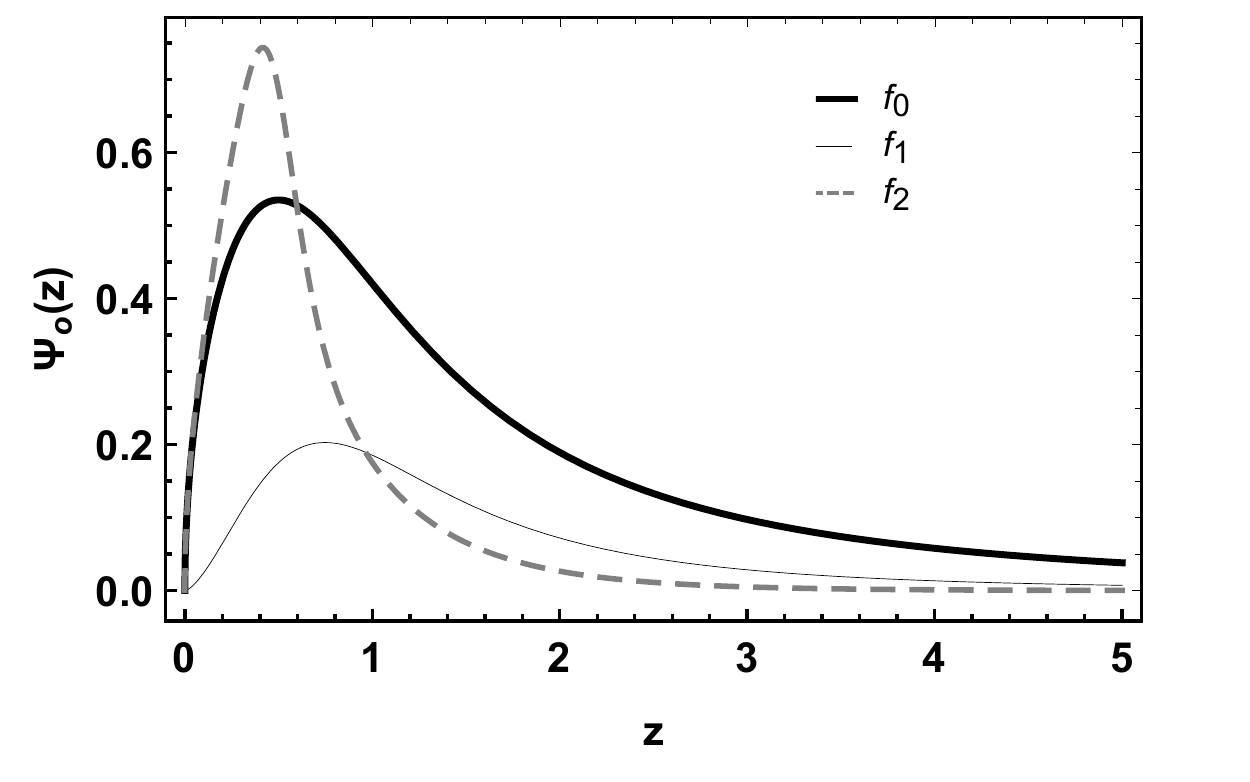}\\
(a) \hspace{8 cm}(b)
\end{tabular}
\end{center}
\caption{(a) Effective potential for $p=\rho=\lambda=1$. (b) Gravitational massless mode for $\rho=\lambda=p=1$.
\label{figPE1}}
\end{figure}

The expression of the potential is too lengthy to be written here. Instead, we plotted the potential in Fig. \ref{figPE1}($a$) ($\beta=0$), in order to explore some qualitative features. Both $f_0$, $f_1$ and $f_2$ are infinite potential well around the origin, as in the GR based string like models \cite{conifold,cigar,regularstring}. For $f_1$ and $f_2$, increasing the value of the parameters $p$, $\lambda$ and $\rho$ increases the amplitude and width of the well. The profile of the massless mode is depicted in Fig. \ref{figPE1}($b$). It is important to emphasize that by increasing the value of the parameters $p$, $\lambda$ and $\rho$ the massless modes become more localized.

\begin{figure}
\begin{center}
\begin{tabular}{ccccccccc}
\includegraphics[height=5cm]{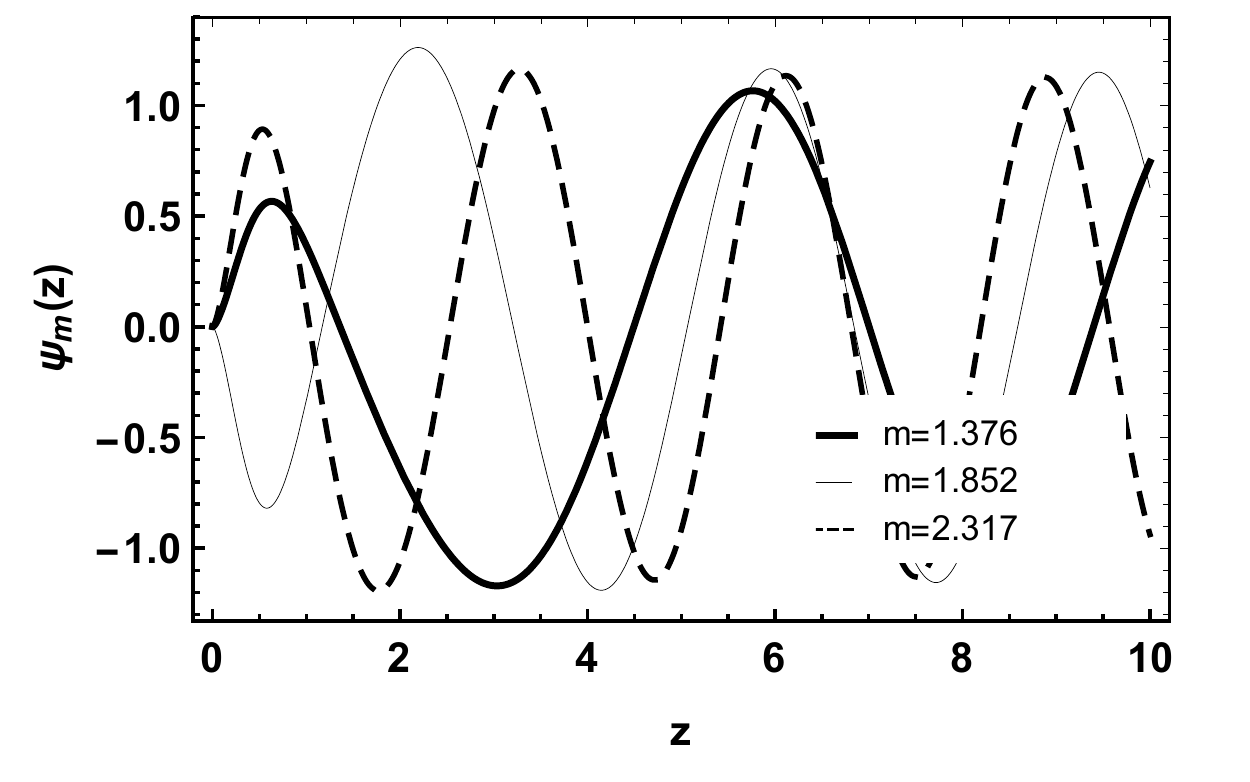}
\includegraphics[height=5cm]{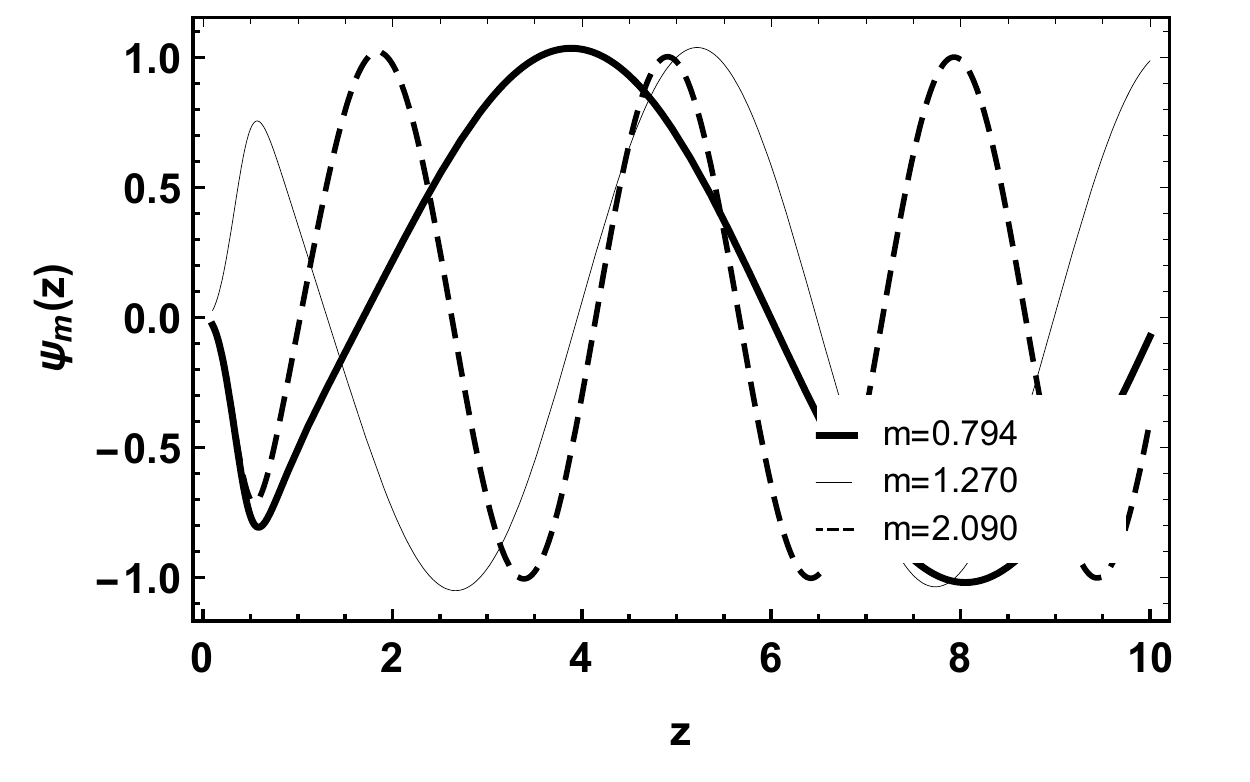}\\
(a) \hspace{8 cm}(b)
\end{tabular}
\end{center}
\caption{Massive modes for $p=\rho=\lambda=1$. (a)$f_1$. (b)$f_2$.
\label{figmasive}}
\end{figure}

We can numerically solve Eq. (\ref{36.l}) and obtain the massive modes. Adopting the usual boundary condition $\psi'(0)=\psi'(\infty)=0$ \cite{Gherghetta,KK1}, the first massive modes are shown in Fig. \ref{figmasive}. Note that in Fig. \ref{figmasive}($a$) for $f_1$, when increasing the mass eigenvalue, there are an increasing of the frequency and the amplitudes of the oscillations are modified around the brane. The first peak close to the origin has a smaller amplitude compared to the others. The same is true for $f_2$, as can be seen in Fig. \ref{figmasive}($b$). Besides, the first peak next to the origin presents a deformation. For $f_0$, when increasing the mass eigenvalue, we note that only the frequency of the oscillations are increased, as expected.

\section{Final remarks}
\label{finalremarks}

We studied the modification of teleparallel gravity on a string-like braneworld. For $f_0(T)=T$, we have the conventional case of teleparallel gravity that is equivalent to GR. For $f_1(T)=T_0\ e^{T/T_0}$ we see the suggestion of a small well in the energy density around the origin. 
For $f_2(T)=T_0 \tanh({T/T_0})$, there are additional peak and well in the energy density around the origin. The additional peaks and wells in the energy density at the brane  suggest the formation of  rings-like structures surrounding the origin.

The effects of the torsion on the KK modes are seen through the analysis of the Schr\"{o}dinger-like potential. The case for $f_0$, as expected, presents an infinite potential well around the origin. Both $f_1$ and $f_2$ cases show an infinite potential barrier around the origin. Right after the barrier, a finite potential well is formed.  We realized then, around the origin the torsion tends to localize the gravitational massless mode in ring-like structures.

Considering massive modes in the $f_0$ case, when the mass eigenvalue increases, it is noted an increasing in the frequency of the oscillations, the same goes for $f_1$ and $f_2$. However, for $f_1$ and $f_2$, by increasing the mass eigenvalue, the amplitude of the oscillations around the brane is modified. This feature is linked to the chosen of the $f(T)$ profiles. Therefore, the torsion that is responsible for the brane splitting process leads to modifications of the massive gravitons inside the thick brane. The first peak close to the origin shows us that the interaction of the massive modes with the torsion in more intense inside the brane core.

\section{Acknowledgments}
The authors thank the Conselho Nacional de Desenvolvimento Cient\'{\i}fico e Tecnol\'{o}gico (CNPq), grants n$\textsuperscript{\underline{\scriptsize o}}$ 312356/2017-0 (JEGS) and n$\textsuperscript{\underline{\scriptsize o}}$ 308638/2015-8 (CASA) and Coordenaçao de Aperfeiçoamento do Pessoal de Nível Superior (CAPES), for financial support.

\end{document}